\documentclass[12pt,a4paper]{article}
\usepackage[pdfstartview=FitH,colorlinks=true,linkcolor=black,anchorcolor=black,citecolor=black,urlcolor=black]{hyperref}
\usepackage[english]{babel}
\usepackage{amsmath,amssymb,titling,authblk}
\usepackage{slashed}
\usepackage{amsmath}
\usepackage{amscd}
\usepackage[normalem]{ulem}
\usepackage{appendix}
\usepackage{bbold}
\usepackage{bm}

\usepackage{pdflscape}
\usepackage{yfonts}
\usepackage{amscd}

\usepackage{cite}

\allowdisplaybreaks[4]
\numberwithin{equation}{section}

\usepackage{color}  

\definecolor{verde}{cmyk}{.83,.21,1,.08}
\definecolor{darkorchid}{rgb}{0.6, 0.2, 0.8}

\newcommand{\be}{\begin{equation}}
\newcommand{\ee}{\end{equation}}
\newcommand{\bea}{\begin{eqnarray}}
\newcommand{\eea}{\end{eqnarray}}
\newcommand{\ii}{\mathrm{i}}

\newcommand{\qed}{\nobreak \ifvmode \relax \else
      \ifdim\lastskip<1.5em \hskip-\lastskip
      \hskip1.5em plus0em minus0.5em \fi \nobreak
      \vrule height0.75em width0.5em depth0.25em\fi}

\newcommand{\gz}{\cdot}

\begin{document}

\setlength{\droptitle}{-6pc}

\title{Clifford Structures in Noncommutative Geometry and\\[2pt] the Extended Scalar Sector\vspace{5pt}}

\renewcommand\Affilfont{\itshape}
\setlength{\affilsep}{1.5em}
\renewcommand\Authands{ and }

\author[1,2,3]{Maxim A.~Kurkov\thanks{max.kurkov@gmail.com}}
\author[2,3,4]{Fedele Lizzi\thanks{fedele.lizzi@na.infn.it}}
\affil[1]{
CMCC-Universidade Federal do ABC, Santo Andr\'e, S.P., Brazil
\vspace{5pt}}
\affil[2]{Dipartimento di Fisica ``Ettore Pancini'', Universit\`{a} di Napoli {\sl Federico II}\vspace{5pt}, Napoli, Italy}
\affil[3]{INFN, Sezione di Napoli, Italy\vspace{5pt}}
\affil[4]{Departament de F\'{\i}sica Qu\`antica i Astrof\'{\i}sica and Institut de C\'{\i}encies del Cosmos (ICCUB),
Universitat de Barcelona. Barcelona, Spain}

\date{}

\maketitle 

\begin{abstract}\noindent
{We consider aspects of the noncommutative approach to the standard model based on the spectral action principle. We show that as a consequence of the incorporation of the Clifford structures in the formalism, the spectral action contains an extended scalar sector, with respect to the minimal Standard Model. This may have interesting phenomenological consequences. Some of these new scalar fields carry both weak isospin and colour indexes. We calculate the new terms in spectral action due to the presence of these fields. 
Our analysis demonstrates that the fermionic doubling in the noncommutative geometry is not just a presence of spurious degrees of freedom, but it is an interesting and peculiar property of the formalism, which leads to physically valuable conclusions. Some of the new  fields do not contribute to the physical fermionic action, but they appear in the bosonic spectral action. Their contributions to the Dirac operator correspond to couplings with the spurious fermions, which are projected out.} 
\end{abstract}
\newpage

\medskip

\section{Introduction}

The standard model of particle interactions can be efficiently described by a particular noncommutative geometry: an ``almost commutative geometry''. Over the years the model has been developing both in its mathematical and physical aspects. Its mathematical framework has its roots in a global view~\cite{Connesbook, Landibook, Ticosbook, ConnesMarcollibook} of geometry based on the spectral properties of operators. The applications of this point of view to geometry are quite startling, the standard reference of the model in its modern version is~\cite{AC2M2}, for a recent review see~\cite{Walterbook}. The model has predictive power, although it is premature to consider it a fully fledged theory to confront with experiment, with prediction with a significative number of digits. Its main success is in the description of the symmetries of the model, very few Yang-Mills models can be described by a noncommutative geometry (NCG), but the standard model and few more can. The Higgs field emerges naturally as an intermediate boson corresponding to the noncommutative part of the model, of a par with photons, $W$, $Z$ and gluons. The actions for fermions and bosons are firmly based on the  spectral properties of a generalized Dirac operator~\cite{SpectralAction}\footnote{{It is remarkable the  spectral action is intimately connected to  anomalies~\cite{AndrianovLizzi, KurkovAndrianovLizzi, KurkovLizziHiggs}, and further development of this observation  leads to interesting results beyond the noncommutative geometry~\cite{MaxMairi}.}}  and the procedure is capable of obtaining numbers such as the mass of the Higgs. The numbers produced in~\cite{AC2M2}, although encouraging, are not in agreement with present data, in particular the model requires the unification of all couplings at a single energy, and one calculates the Higgs boson mass around 170~GeV. Both these aspects are experimentally excluded, and the model can be fixed to allow the physical mass of the Higgs boson~\cite{Resilience, Grand1, Grand2, ChamsConnesWalter1,ChamsConnesWalter2, AgostinoPierre, ChamsConnesWalter3, Aydemir:2013zua, Aydemir:2014ama, Minic}. Efforts are also undertaken to use the model for other predictions, for example in~\cite{Aydemir:2016xtj}. 

The fact that the calculations made in the present model are encouraging, but not yet comparable with experiment, suggests that some improvement may happen also from the mathematical side. In~\cite{FrancescoLudwik} it was discussed a noncommutative version of the Clifford symmetry. One of the remarkable effects of the Clifford requirements is the appearance of scalar fields which are not present in the usual description.

The aim of this paper is to discuss in detail these new fields and their couplings.  In particular we will calculate their contribution to the spectral action. The noncommutative model is by nature Euclidean and exhibits spurious degrees of freedom, known as ``fermion doubling''~\cite{LMMS}, therefore for physical applications a Wick (anti)rotation accompanied by an elimination of these spuriuous degrees of freedom is necessary. We have described this procedure in detail in~\cite{direstraits}. Here we find that not all of these extra bosons behave upon this procedure in a standard way: some of the new scalar fields present in the Euclidean Dirac operator are absent in the corresponding (Lorentzian) physical action for fermions.

The paper is organized as follows: in Sec.~\ref{revsetup} we review the noncommutative geometric approach to the Standard Model, focusing on the modification of the formalism due to an introduction of the Clifford structures proposed in~\cite{FrancescoLudwik}. In Sec.~\ref{flu} we introduce the new scalar fields, which come out from the fluctuations of the Dirac operator in the ``Clifford-based" approach~\cite{FrancescoLudwik}, and discuss their transformation properties upon the action of the gauge group. Sec.~\ref{BSA} is devoted to the bosonic spectral action: we compute the new terms with respect to the ``standard" spectral approach~\cite{AC2M2}. In Sec.~\ref{physact} we discuss the physical action derived from this model: we carry out the Wick rotation to the Lorentzian signature and get rid of the spurious degrees of freedom in the fermionic action.
The last section contains our conclusions.

\section{The standard model as a Noncommutative Geometry \label{revsetup}}

In this section we sketch the main aspects of the model. We will be very brief, the reader familiar with this approach will need this section just to set the notations. First we outline the basic concepts of the spectral triples, which are common for both the ``standard" approach~\cite{AC2M2} and the ``Clifford-based"~\cite{FrancescoLudwik} approaches, whilst afterwords we discuss the peculiar features of the latter, which differ it from the former: the finite dimensional grading $\gamma_F$ and the finite dimensional Dirac operator $D_F$.

\subsection{The Standard Spectral Triple}

In the spectral approach a geometry is described by a \emph{spectral triple}~\cite{Connesbook, Landibook, Ticosbook}, i.e.\ a $*$-algebra (possibly noncommutative) realized as bounded operators on a Hilbert space, and a self adjoint operator which generalizes the Dirac operator. The algebra describes the topology of the space, for the case at hand the Hilbert space describes the matter content and the Dirac operator gives a metric structure and enables the writing of action. Being based on operators all quantities are based on spectra, and in particular the actions for bosons and fermions can be written in purely spectral form. Also of fundamental importance are  two more operators: the grading and the real structure,  which generalize chirality (for the even dimensional case) and charge conjugation.  The standard model emerges form this scheme. We will briefly describe this approach mainly to set notations, referring for details to the original literature~\cite{SpectralAction, AC2M2} or the recent book~\cite{Walterbook}.
We start choosing an algebra which is the product commutative infinite dimensional algebra of continuous functions on the manifold $\mathcal{M}$, which represents the space-time  times a noncommutative but finite dimensional matrix algebra
\be
\mathcal A=C(\mathcal{M})\otimes \mathcal A_F
\ee
For the standard model the finite algebra is 
\be\mathcal A_F=\mathbb C\oplus\mathbb H\oplus \mathrm{Mat}_3(\mathbb C) \label{algebraF}
\ee 
where by $\mathbb H$ we indicate quaternions, and by $\mathrm{Mat}_3(\mathbb C)$ three by three complex matrices.
Likewise the Hilbert space is the product of usual spinors times a finite dimensional Hilbert space, which contains all physical degrees of freedom:
\be
\mathcal H=\mathrm{sp}(\mathcal{M})\otimes \mathcal H_F \label{HilbertSpace}
\ee
the generalized Dirac operator (which in the following we will simply call Dirac operator) is
\be
\mathcal D_0=  \ii \gamma^\mu \nabla^{\mathrm{LC}}_\mu\otimes 1_F + \gamma^5 \otimes {D}_F \label{D0}
\ee
Where $\nabla_\mu^{\mathrm{LC}}$ is the covariant derivative on the spinor bundle of $\mathcal{M}$, which contains the Levi-Civita spin connection. Gravity in the action is considered background, and is not quantized. A curved background does not however play a major role in this paper, but is useful to retain it, as it enables some simplification in the calculations, as we will see in Sect.~\ref{physact}.

As we mentioned, there are two more operators which play an important role. They are the grading operator $\Gamma$ and the antiunitary real structure $\mathcal{J}$.  The grading operator $\Gamma$ is present in the even dimensional case, it satisfies $\Gamma^2=\mathbb 1$ and  it is  taken to be
\be
\Gamma=\gamma^5\otimes \gamma_F
\ee
where $\gamma^5$ is the chirality matrix i.e. the usual  product of all four Dirac's $\gamma^\mu$ and $\gamma_F$ is an operator acting on $\mathcal H_F$. It is usually taken to have eigenvalue +1 on left handed states, and -1 on right handed one, but other choices are possible and we will discuss them later in the paper.

The real structure operator $\mathcal J=J\otimes J_F$, which is \emph{antiunitary}  in $\mathcal H$, enables the definition of the opposite algebra 
\be
\mathcal A^o=\mathcal{JAJ}^{-1}.
\ee
The elements of the triple must satisfy several conditions, which render the space the noncommutative equivalent of a manifold~\cite{Connesmanifold}. There are conditions of compatibility between $\Gamma, \mathcal J$ and $\mathcal D_0$ with signs which depend on the dimensions:
\be
\mathcal J^2=\pm \mathbb 1 \ , \ \mathcal J \Gamma=\pm\Gamma\mathcal J \ ,\  \mathcal J \mathcal D_0=\pm \mathcal D_0\mathcal J 
\ee
 The opposite algebra must commute with the algebra (order zero condition):
\be
[a,\mathcal J b \mathcal J^{-1}]=0 \ , \ \forall \, a,b\in\mathcal A \label{orderzero}
\ee 
and with one forms (the order one condition)
\be
[[\mathcal D_0,a],\mathcal J b \mathcal J^{-1}]=0 \ , \ \forall \, a,b\in\mathcal A \label{orderone}
\ee

The dimension of $\mathcal H_F$ in~\eqref{HilbertSpace} is 96. This number is obtained taking into account that there is a lepton left doublet plus two right handed singlets, and a doublet and two singlets for quarks times three colours. This makes 16 degrees of freedom, times 3 generations, and times two for particle/antiparticle, sums to 96. Since the spinor index has four degrees of freedom the element of the full Hilbert space $\mathcal{H}$ is described by  384 independent complex valued functions. Clearly there is some overcounting, called for historical reasons \emph{fermion doubling}~\cite{LMMS}. We will come back to this issue, as well as the fact that the model is at this stage Euclidean, in section~\ref{physact}.

We will label the elements of $\mathcal H_F$ according the basis given by the elementary particles of the standard model (including right handed neutrinos):
\be
( \boldsymbol \nu_R, \boldsymbol e_R, \boldsymbol L_L,\boldsymbol u_R, \boldsymbol{d_R}, \boldsymbol Q_L, \boldsymbol \nu_R^c, \boldsymbol e_R^c, \boldsymbol L_L^c,
\boldsymbol u_R^c, \boldsymbol{d_R^c}, \boldsymbol Q_L^c ) \label{HfStruct}
\ee
where $Q_L$ corresponds to\footnote{The construction of the product space clarifies in which sense the word ``corresponds" is used: see in particular the discussion around \eqref{HStruct}.} the quark doublet $(\boldsymbol u_L, \boldsymbol  d_L)$ while $L_L$ corresponds to the lepton doublet $(\boldsymbol \nu_L, \boldsymbol e_L)$, with the supercript $c$  we indicate the elements of $H_F$ which correspond to the antiparticles and by boldface characters we indicate that the elements have to replicated by three generations, for example $\boldsymbol e=(e,\mu,\tau)$ and so on.
Quarks have an extra colour index, which we omit. 
Below we will use the following notation for matrices action on $\mathcal H_F$. We define the matrix unity $E_{\boldsymbol u_R \boldsymbol u_R}$ to be a matrix whose only nonzero element is an identity matrix in the ${\boldsymbol u_R}$ location, likewise for $E_{\boldsymbol u_R \boldsymbol d_R}$ is an off diagonal matrix with nonvanishing entry in the ${\boldsymbol u_R \boldsymbol d_R}$, and so on. In the cases for which a singlet crosses a doublet then we assume that, for example, $E_{\boldsymbol u_R,\boldsymbol L_L}$ is two identity matrices side by side, or vertically superimposed. 

The representation of the algebra is diagonal and with our notation, an element $a=(\lambda,h,m)$ with  $\lambda\in\mathbb C, h\in\mathbb H$ and $m\in\mbox{Mat}_3(\mathbb C)$ is represented by the matrix\footnote{Here and in the following we omit terms like $\otimes\mathbb 1_3$ when for example 
a complex number act on a quark, and likewise for doublets etc.} :
\bea
a&=& \lambda E_{\boldsymbol u_R,\boldsymbol u_R} +\lambda^* E_{\boldsymbol d_R,\boldsymbol d_R} + h E_{\boldsymbol Q_L,\boldsymbol Q_L} + \lambda E_{\boldsymbol \nu_R,\boldsymbol \nu_R}+ \lambda^* E_{ \boldsymbol e_R,\boldsymbol e_R} +h E_{\boldsymbol L_L,\boldsymbol L_L}+\nonumber\\ &&m E_{\boldsymbol u_R^c,\boldsymbol u_R^c} + m E_{\boldsymbol d_R^c, \boldsymbol d_R^c} + m E_{\boldsymbol Q_L^c,\boldsymbol Q_L^c} +\lambda E_{\boldsymbol \nu_R^c,\boldsymbol \nu_R^c}  +\lambda E_{ \boldsymbol e_R^c,\boldsymbol e_R^c} + \lambda E_{\boldsymbol L_L^c,\boldsymbol L_L^c}.\nonumber \label{algebraconproiettori}
\eea
In our notations the real structure $J_F$ of the finite spectral triple reads:
\be
J_F=( E_{\boldsymbol u_R,\boldsymbol u_R^c} +E_{\boldsymbol d_R,\boldsymbol d_R^c} +E_{\boldsymbol Q_L,\boldsymbol Q_L^c} +E_{\boldsymbol \nu_R,\boldsymbol \nu_R^c} + E_{ \boldsymbol e_R,\boldsymbol e_R^c} +E_{\boldsymbol L_L,\boldsymbol L_L}) cc
\ee
where $cc$ is complex conjugation.

So far we have been in the framework of~\cite{AC2M2}. From now on we focus on the peculiar properties of the construction of ~\cite{FrancescoLudwik}, which enables to incorporate the Clifford structures in the finite spectral triple. We refer to the original paper for all the details, and present here just the results.

\subsection{Alternative Grading}
The first novelty of the Clifford based construction is the grading $\gamma_F$ of the finite spectral triple, which has the following form:
\bea
\gamma_F &=&- E_{\boldsymbol u_R,\boldsymbol u_R} -E_{\boldsymbol d_R,\boldsymbol d_R} +E_{\boldsymbol Q_L,\boldsymbol Q_L} 
+ E_{\boldsymbol \nu_R,\boldsymbol \nu_R} +E_{ \boldsymbol e_R,\boldsymbol e_R} -E_{\boldsymbol L_L,\boldsymbol L_L}\nonumber\\ &&- (- E_{\boldsymbol u_R^c,\boldsymbol u_R^c} -E_{\boldsymbol d_R^c, \boldsymbol d_R^c} +E_{\boldsymbol Q_L^c,\boldsymbol Q_L^c} +E_{\boldsymbol \nu_R^c,\boldsymbol \nu_R^c} +E_{ \boldsymbol e_R^c,\boldsymbol e_R^c} -E_{\boldsymbol L_L^c,\boldsymbol L_L^c}),
\eea
which differs from the ``standard grading"  $\gamma^{\mathrm{st}}_F$ considered in \cite{AC2M2}: 
\bea
\gamma^{\mathrm{st}}_F&=&- E_{\boldsymbol u_R,\boldsymbol u_R} -E_{\boldsymbol d_R,\boldsymbol d_R} +E_{\boldsymbol Q_L,\boldsymbol Q_L} -E_{\boldsymbol \nu_R,\boldsymbol \nu_R} -E_{ \boldsymbol e_R,\boldsymbol e_R} +E_{\boldsymbol L_L,\boldsymbol L_L}\nonumber\\ &&- (- E_{\boldsymbol u_R^c,\boldsymbol u_R^c} -E_{\boldsymbol d_R^c, \boldsymbol d_R^c} +E_{\boldsymbol Q_L^c,\boldsymbol Q_L^c} -E_{\boldsymbol \nu_R^c,\boldsymbol \nu_R^c} -E_{ \boldsymbol e_R^c,\boldsymbol e_R^c} +E_{\boldsymbol L_L^c,\boldsymbol L_L^c}).
\eea
The two are connected by the following formula
\be
\gamma_F = \left(\mathrm{Q} - \mathrm{L} \right)\gamma^{\mathrm{st}}_F,
\ee
where $\mathrm{Q}$ and $\mathrm{L}$ stand for the projectors of the ``quark" and ``leptonic" subspaces of $\mathcal{H}_F$ respectively:
\bea
\mathrm{Q} &=&  E_{\boldsymbol u_R,\boldsymbol u_R} + E_{\boldsymbol d_R,\boldsymbol d_R} +E_{\boldsymbol Q_L,\boldsymbol Q_L} 
+ E_{\boldsymbol u_R^c,\boldsymbol u_R^c} + E_{\boldsymbol d_R^c, \boldsymbol d_R^c} +E_{\boldsymbol Q_L^c,\boldsymbol Q_L^c},
\nonumber\\
\mathrm{L} &=& E_{\boldsymbol \nu_R,\boldsymbol \nu_R} + E_{ \boldsymbol e_R,\boldsymbol e_R} +E_{\boldsymbol L_L,\boldsymbol L_L} + E_{\boldsymbol \nu_R^c,\boldsymbol \nu_R^c} + E_{ \boldsymbol e_R^c,\boldsymbol e_R^c} +E_{\boldsymbol L_L^c,\boldsymbol L_L^c}.
\eea

\subsection{The Dirac Operator}
Another novelty of the Clifford based approach is the Dirac operator $D_F$, which has the following form:
\bea
D_F &=&\boldsymbol\Upsilon_{\boldsymbol \nu} E_{\boldsymbol \nu_R \boldsymbol L_L}
+\boldsymbol\Upsilon_{\boldsymbol e} E_{\boldsymbol e_R \boldsymbol L_L}
+\boldsymbol\Upsilon_{\boldsymbol u} E_{\boldsymbol u_R \boldsymbol Q_L}
+\boldsymbol\Upsilon_{\boldsymbol d} E_{\boldsymbol d_R \boldsymbol Q_L}\nonumber\\
&&+ {\boldsymbol\Omega}^* E_{\boldsymbol \nu_R \boldsymbol e_R^c} +\boldsymbol \Delta_U E_{\boldsymbol \nu_R^c u_R^c}+\boldsymbol \Delta_D E_{\boldsymbol e_R^c d_R^c}+\boldsymbol \Delta_L E_{\boldsymbol L_L^c Q_L^c} +\boldsymbol {\mathrm K}E_{\boldsymbol L_L \boldsymbol u_R^c}   \nonumber\\
&&+ J_F \left(\boldsymbol\Upsilon_{\boldsymbol \nu} E_{\boldsymbol \nu_R \boldsymbol L_L}
+\boldsymbol\Upsilon_{\boldsymbol e} E_{\boldsymbol e_R \boldsymbol L_L}
+\boldsymbol\Upsilon_{\boldsymbol u} E_{\boldsymbol u_R \boldsymbol Q_L}
+\boldsymbol\Upsilon_{\boldsymbol d} E_{\boldsymbol d_R \boldsymbol Q_L}\right. \nonumber\\
&& \left.+{\boldsymbol\Omega}^* E_{\boldsymbol \nu_R \boldsymbol e_R^c} +\boldsymbol \Delta_U E_{\boldsymbol \nu_R^c u_R^c}+\boldsymbol \Delta_D E_{\boldsymbol e_R^c d_R^c}+\boldsymbol \Delta_L E_{\boldsymbol L_L^c Q_L^c} +\boldsymbol {\mathrm K}E_{\boldsymbol L_L \boldsymbol u_R^c}\right)J_F \nonumber\\
&&+ {\boldsymbol\Upsilon}_R^{\dagger} E_{\boldsymbol \nu_R\boldsymbol\nu_R^c}
+h.c.
, \label{D0}
\eea
and which is compatible with the new grading $\gamma_F$ and other requirements of the approach of~\cite{FrancescoLudwik}. 
 The terms on the first on the third and on the last lines involve the usual Yukawa couplings and the Majorana mass terms, which are already present in \cite{AC2M2}. The second and the fourth lines instead contain novel terms, which are the object of this paper: $\boldsymbol \Delta$ and $\boldsymbol{\mathrm K}$ provide novel couplings of leptons and quarks, ${\boldsymbol\Omega}$ couples leptons among themselves in the Euclidean action \emph{before the projection on the physical subspace}. We will see later on that the projection to the physical subspace will eliminate some of these couplings. It is important that the selfconsistency the approach of~\cite{FrancescoLudwik} requires in particular that: 
\begin{itemize}
\item{both entries $\boldsymbol \Delta_D$ and $\boldsymbol \Delta_L$ must differ from zero},
\item{and at least two out of the three entries $\boldsymbol \Delta_U$, $\boldsymbol{\mathrm K}$ and ${\boldsymbol\Omega}$ must be different from zero.}
\end{itemize}

In conclusion we present the explicit matrix form of the Dirac operator $D_F$ defined by \eqref{D0}:
\begingroup
\setlength{\arraycolsep}{11.5pt}
\renewcommand{\arraystretch}{1.2}
\be
D_F=\left[\!\begin{array}{cccccc|cccccc}
\gz & \gz &\boldsymbol  \Upsilon_\nu\!\!\! & {\boldsymbol \Delta_u}^*\!\!\! & \gz & \gz & \boldsymbol {{\Upsilon}}^{\dagger}_R\!\!\! &\boldsymbol { {\Omega}}^* & \gz & \gz & \gz & \gz \\
\gz & \gz &\boldsymbol  \Upsilon_e\!\!\! & \gz & \boldsymbol {\Delta_d}^*\!\!\! & \gz & \boldsymbol {{\Omega}}^{\dagger} & \gz & \gz & \gz & \gz & \gz \\
\boldsymbol \Upsilon_\nu^\dag\!\!\! & \boldsymbol \Upsilon_e^\dag\!\!\! & \gz & \gz & \gz & \boldsymbol {\Delta}^*_L\!\!\! & \gz & \gz & \gz & \boldsymbol {\mathrm{K}} & \gz & \gz \\
\boldsymbol \Delta_u^t\!\!\! & \gz & \gz & \gz & \gz & \boldsymbol \Upsilon_u\!\!\! & \gz & \gz & \boldsymbol {\mathrm{K}}^t\!\! & \gz & \gz & \gz \\
\gz &\boldsymbol  \Delta_d^t\!\!\! & \gz & \gz & \gz & \boldsymbol \Upsilon_d\!\!\! & \gz & \gz & \gz & \gz & \gz & \gz \\
\gz & \gz &\boldsymbol  \Delta_L^t\!\!\! &\boldsymbol  \Upsilon_u^\dag\!\!\! &\boldsymbol  \Upsilon_d^\dag\!\!\! & \gz & \gz & \gz & \gz & \gz & \gz & \gz \\
\hline
\boldsymbol \Upsilon_R\!\!\! & \boldsymbol \Omega & \gz & \gz & \gz & \gz & \gz & \gz & \boldsymbol {\Upsilon}^*_\nu\!\!\! &\boldsymbol  \Delta_u\!\!\! & \gz & \gz \\
\boldsymbol \Omega^{t} & \gz & \gz & \gz & \gz & \gz & \gz & \gz &\boldsymbol{ \Upsilon}^*_e\!\!\! & \gz & \boldsymbol \Delta_d\!\!\! & \gz \\
\gz & \gz & \gz & {\boldsymbol {\mathrm{K}^*}} & \gz & \gz &\boldsymbol  \Upsilon_\nu^t\!\!\! & \boldsymbol \Upsilon_e^t\!\!\! & \gz & \gz & \gz &\boldsymbol  \Delta_L\!\!\! \\
\gz & \gz & \boldsymbol {\mathrm{K}}^\dag\!\! & \gz & \gz & \gz & \boldsymbol \Delta_u^\dag\!\!\! & \gz & \gz & \gz & \gz & 
{\boldsymbol \Upsilon}^*_u \\
\gz & \gz & \gz & \gz & \gz & \gz & \gz &\boldsymbol  \Delta_d^\dag\!\!\! & \gz & \gz & \gz & {\boldsymbol \Upsilon}^*_d \\
\gz & \gz & \gz & \gz & \gz & \gz & \gz & \gz &\boldsymbol  \Delta_L^\dag\!\!\! &\boldsymbol  \Upsilon_u^t\!\!\! & \boldsymbol \Upsilon_d^t\!\!\! & \gz
\end{array}\!\right].  \label{DFbig}
\ee
\endgroup
Setting $\boldsymbol  \Delta_{U,D,L} = 0$, $\boldsymbol \Omega = 0$ and $\boldsymbol K = 0$, one obtains the standard $D_F$ of~\cite{AC2M2}.

\section{Fluctuations of the Dirac operator: Fields \label{flu}}
The fluctuated Dirac operator is constructed in the following way:
\be
\mathcal{D} = \mathcal D_0 +\sum_i a_i [\mathcal D_0,b_i] +\sum_i  \mathcal J a_i  [\mathcal D_0,b_i] \mathcal J^\dagger   \label{fluctu},
\ee
for generic elements $a_i, b_i\in\mathcal A$. Both gauge and scalar fields in the spectral approach come out from these fluctuations.
Presence of the new terms (with respect to \cite{AC2M2}) in \eqref{DFbig} indicates new scalar fields, not present in the Standard Model.

Below we restrict ourselves to the following structures, where the dependence on the generation indexes is factorised:
\bea
\boldsymbol  \Upsilon_{\nu} &=& \hat{Y}_{u} \otimes \tilde{h}_{\nu}^{\dagger} \nonumber\\
\boldsymbol  \Upsilon_{e} &=& \hat{Y}_{d} \otimes h_{e}^{\dagger} \nonumber\\
\boldsymbol  \Upsilon_{u} &=&  \hat{y}_{u} \otimes \tilde{h}_{u}^{\dagger} \nonumber \\
\boldsymbol  \Upsilon_{d} &=&  \hat{y}_{d} \otimes {h}_d^{\dagger}\nonumber\\
\boldsymbol {\Delta_u}^* &=& \hat{y}_{\Delta_u}^{\dagger}\otimes d_u^{\dagger}\nonumber \\
\boldsymbol {\Delta_d}^* &=& \hat{y}_{\Delta_d}^{\dagger}\otimes d_d^{\dagger}\nonumber \\
\boldsymbol {\Delta_L}^* &=& \hat{y}_{\Delta_L}^{\dagger}\otimes d_L^{\dagger}\nonumber \\
\boldsymbol {\mathrm{K}}  &=& \hat{y}_{S}^{\dagger}\otimes s^{\dagger}, \nonumber\\
\boldsymbol {{\mathrm{\Omega^*}}}  &=& \hat{y}_{\Omega}\otimes \omega, \nonumber\\
\boldsymbol {\Upsilon_R^{\dagger}} &=& \hat{y}_{M}\otimes\mathrm{M}_R.  \label{choice}
\eea
In these formulas the two component columns $h_{\nu,e,u,d}$ (in the Weak isospin indexes) are chosen in the same way as it was done in~\cite{AC2M2} (hereafter $v$ is an arbitrary complex constant of the dimension of the mass): 
\be
h_{\nu} = \left(\begin{array}{c} v \\ 0   \end{array}\right), 
\quad h_{e} = \left(\begin{array}{c} 0 \\ v   \end{array}\right),
\quad h_{u} = \left(\begin{array}{c} v \\ 0   \end{array}\right), 
\quad h_{d} = \left(\begin{array}{c} 0 \\ v   \end{array}\right),
\ee
and the three component columns $d_{u,d,L}$ (in the colour indexes) we choose as follows:
\be
d_{u} = \left(\begin{array}{c} v \\ 0   \\ 0 \end{array}\right), 
\quad d_{d} = \left(\begin{array}{c} 0 \\ v  \\ 0  \end{array}\right),
\quad d_{L} = \left(\begin{array}{c} 0 \\ 0   \\ v \end{array}\right). 
\ee
The quantity  $s$ is the complex 3 by 2 matrix (in both colour and the weak isospin indexes):
\be
s = \left(\begin{array}{cc} v & 0 \\ 0 & 0 \\ 0 & 0  \end{array}\right),
\ee
 $\omega$ is the complex number, which we set to $v$, the dimensionful constant  $\mathrm{M}_R$ sets the Majorana mass scale for the right handed neutrinos, which is needed for the sea-saw mechanism.  The quantities $\hat{Y}_{u}$, $\hat{Y}_{d}$, $\hat{y}_{u}$, $\hat{y}_{d}$, 
$\hat{y}_{\Delta_u}$, $\hat{y}_{\Delta_d}$, $\hat{y}_{\Delta_L}$, 
$\hat{y}_{S}$ and $ \hat{y}_{M}$ are arbitrary (dimensionless) complex 3 by 3 Yukawa matrices which act on the generation index. The tilde indicates charge conjugated weak isospin doublets e.g. $\tilde{h}_{\nu} = \sigma_2 {h}^*_{\nu}$, where $\sigma_2$ stands for the second Pauli matrix.

Considering the fluctuations  \eqref{fluctu} of the Dirac operator one can see, that in order to construct the fluctuated Dirac operator, $\mathcal{D}$ one has replace the constant matrices in \eqref{choice} by the matrix valued functions according to the following rule:
\bea
\tilde h_{\nu} &\longrightarrow& \tilde{H} \nonumber\\
 h_{e}   &\longrightarrow&  H \nonumber \\
\tilde  h_{u}       &\longrightarrow& \tilde{H} \nonumber\\
   h_{d}     &\longrightarrow& H \nonumber\\
d_u &\longrightarrow& \Delta_u \nonumber \\
d_d &\longrightarrow& \Delta_d \nonumber \\
d_L &\longrightarrow& \Delta_L \nonumber \\
s &\longrightarrow& S \nonumber\\
\omega &\longrightarrow& \Omega 
  \label{rules}.
\eea
Note that upon the fluctuations of the Dirac operator $\mathrm{M}_R$ remains a constant i.e. it does not transform 
into a field.

By definition the gauge subgroups $SU(2)$ and $SU(3)$ are represented 
on the weak isospin fermionic doublets and colour fermionic triplets as a left multiplication by the
unitary matrices $U_{SU(2)}$ and $U_{SU(3)}$ respectively\footnote{We assume that the the components of the weak isospin fermionic doublets and the color fermionic triplets are combined into  \emph{columns}.
Note that antiquarks
and antileptons are transformed by the {complex}  conjugated matrices.}:
\begin{equation}
\left[\mbox{ferm. doublet}\right] 
\rightarrow 
U_{SU(2)} \cdot \left[\mbox{ferm. doublet}\right]; 
\quad \left[\mbox{ferm. triplet}\right] \rightarrow U_{SU(3)} \cdot \left[\mbox{ferm. triplet}\right],
\end{equation}
while the gauge fields transform upon the adjoint representation of the gauge group. The transformation law of the scalar fields which is presented below maintains the gauge invariance of the
fermionic action upon the simultaneous gauge transformation of the fermionic multiplets, gauge and scalar fields.
In what follows $Y$ stands for the abelian hypercharge of a given multiplet, which describes the action of the $U(1)$ gauge subgroup. 

The scalar doublet $H$ is nothing but the Higgs field of the minimal Standard Model, which transforms as follows:
\begin{equation}
H = \left[
\begin{array}{c}
H^{\mathrm{up}} \\ H^{\mathrm{down}}
\end{array}
\right] \begin{CD}@>>{ SU(2)\times SU(3)}>\end{CD} U_{SU(2)} \cdot H; \quad {\rm Y}_H = 1.
\end{equation}
The field $\tilde{H}$ 
transforms as $H$ under the $SU(2)$ transformations however it has the opposite hypercharge:
\begin{equation}
\tilde H = \left[
\begin{array}{c}
\left(H^{\mathrm{down}}\right)^* \\ -\left( H^{\mathrm{up}}\right)^*
\end{array}
\right] \begin{CD}@>>{ SU(2)\times SU(3)}>\end{CD} U_{SU(2)} \cdot \tilde H; \quad {\rm Y}_{\tilde H} = -1.
\end{equation}
For each of the three fields $\Delta_u$, $\Delta_d$ and $\Delta_L$ the transformation law reads:
\begin{equation}
\Delta_{u,d,L} = \left[
\begin{array}{c}
\Delta^{\mathrm{red}}_{u,d,L} \\ \Delta^{\mathrm{green}}_{u,d,L} \\ \Delta^{\mathrm{blue}}_{u,d,L}
\end{array}
\right] \begin{CD}@>>{ SU(2)\times SU(3)}>\end{CD}
 U_{SU(3)} \cdot \Delta_{u,d,L}; 
\quad {\rm Y}_{\Delta_u} = {\rm Y}_{\Delta_d}  = {\rm Y}_{\Delta_L} = \frac{4}{3}.
\end{equation}
The field $S$ carries both colour and weak isospin indexes and transforms in the following way:
\begin{equation}
S = \left[
\begin{array}{cc}
S^{\mathrm{up}\, \mathrm{red}} & S^{\mathrm{down} \, \mathrm{red}} \\ S^{\mathrm{up} \, \mathrm{green}} & S^{\mathrm{down} \, \mathrm{green}} \\ S^{\mathrm{up} \, \mathrm{blue}} &  S^{\mathrm{down} \, \mathrm{blue}}
\end{array}
\right] \begin{CD}@>>{ SU(2)\times SU(3)}>\end{CD}  
U_{SU(3)}^{-1{\rm T}} \cdot S \cdot U_{SU(2)}^{-1}, \quad {\rm Y}_S = { - \frac{1}{3}}.
\end{equation}
The last field $\Omega$ is the $SU(2)\times SU(3)$ singlet, and it transforms nontrivially just under the  $U(1)$ transformations:
\begin{equation}
\Omega  \begin{CD}@>>{ SU(2)\times SU(3)}>\end{CD}  
\Omega, \quad {\rm Y}_\Omega = { - 2}.
\end{equation} 
In the next section we compute the bosonic spectral action.

\section{Bosonic Spectral Action \label{BSA}}
The aim of this section is to calculate the bosonic spectral action\footnote{In the present paper we exploit the ``standard" definition of the bosonic spectral action, which is based on the introduction of the ultraviolet cutoff. Other definitions, are based e.g. on the $\zeta$-function regularisation are also possible \cite{zeta}. }
\begin{equation}
S_B \equiv {\rm Tr }\, \chi \left(\frac{\mathcal{D}^2}{\Lambda^2}\right) \simeq \Lambda^4 f_0 a_0 +\Lambda^2 f_2 a_2 
+ \Lambda^0 f_4 a_4 + O\left(\frac{1}{\Lambda^2}\right), \label{anz}
\end{equation}
where $\chi$ is some cutoff function, $f_0$, $f_2$, $f_4$ are the first three momenta of its Fuorier transform
and $a_0$, $a_2$ and $a_4$ are the first three nontrivial heat kernel coefficients on the manifold without
boundary. The ``fluctuated" (or covariant) Dirac operator is given by:
\be
\mathcal{D} =\ii\gamma^{\mu}  \nabla_{\mu}  + \gamma_{5}\otimes M,
\ee
where the covariant derivative $\nabla_{\mu}$ involves the gauge and the Levi-Civita spin connections, whilst the 96 by 96 matrix $M$ is nothing but the ``fluctuated" version of $D_F$, which is obtained from \eqref{DFbig} via the prescription  \eqref{rules}. \\

\noindent{\bf{\emph{{Comment}}}:} {\small We notice that the asymptotic expansion \eqref{anz} correctly describes the behaviour of the trace in the left hand side of \eqref{anz} at the energies below the cutoff scale $\Lambda$, whilst the high momenta behaviour of the bosonic spectral action is drastically different~\cite{KuLiVa}: high momenta bosons do not propagate, see also \cite{Sauersig}. Physically it means that this model becomes strongly coupled at the energies above $\Lambda$ in both $U(1)$, $SU(2)$ and  $SU(3)$ sectors.  A similar high energy phase transition has been considered beyond the scope of the noncommutative geometry,  see e.g.~\cite{Rubakov,ULP}. In what follows we do not discuss the high momenta regime and the mentioned above effects, so from now on the ansatz in the right hand side of \eqref{anz} is identified with the definition of the bosonic spectral action. }\\

We emphasise that the gauge content of these formalism is identical to the one of \cite{AC2M2}, therefore if one sets  $\Delta_{u,d,L} = 0$, $\Omega = 0$ and $S = 0$ our operator $\mathcal{D}$ will coincide with the one of \cite{AC2M2}, hence it is sufficient to calculate the difference
\be
S_B -S_B\big|_{\Delta_{u,d,L} = 0, \Omega = 0, S = 0}.
\ee

\subsection{Computational simplifications}
The structure of the heat kernel coefficients on manifolds without  boundaries is very well known (see e.g.~\cite{manual}), and one can easily see that the scalar fields can contribute
 to $a_2$ through the combination:
\begin{equation}
a_2^{\rm contrib} =\frac{1}{16\pi^2} \int d^4x\,\sqrt{g}\,{\rm tr}\left( E\right) \label{a2}
\end{equation}
and to $a_4$ through the combination:
\begin{equation}
a_4^{\rm contrib} = \frac{1}{16\pi^2}\,\frac{1}{360}\,\int d^4x\,\sqrt{g}\, {\rm tr}\left(-60 R E + 180 E^2\right), \label{a4cg}
\end{equation}
where  by definition 
\be
E \equiv -D^2 - \nabla^2
\ee  
and $R$ stands for a scalar curvature.

Note that the $a_2$ contribution can not contain covariant derivatives of the scalar field: the simplest scalar contribution which involves the scalar fields and their covariant derivatives has the the canonical dimension 3, whilst the integrand in \eqref{a2} must have the canonical dimension~2.  Therefore, to compute $a_2^{\rm contrib}$ is sufficient to neglect the dependence of scalars on coordinates.

Now let us focus on the $a_4$ contribution. The computation of the scalar contribution to $a_4$
drastically simplifies, when the Dirac operator transforms in a homogeneous way upon the local Weyl transformation of the metric tensor and of the scalar fields. 
Even though the Dirac operator $\mathcal{D}$ does not exhibit this property (since it contains the constant Majorana mass terms for the right handed neutrinos) one can write:
\be
{\rm Tr }\,  \chi \left(\frac{\mathcal{D}^2}{\Lambda^2}\right) = {\rm Tr }\, \chi \left(\frac{{\tilde{\mathcal{D}}}^2}{\Lambda^2}\right)\bigg|_{\sigma = \mathrm{M}_R}, 
\ee 
where the ``intermediate" Dirac operator $\tilde{\mathcal{D}}$ is obtained from $\mathcal{D}$ via the replacement of the constant $\mathrm{M}_R$ by the scalar field 
$\sigma$.
This field has no gauge indexes and it has already been considered in the context of the model
to fix the Higgs mass in~\cite{Resilience}.  We emphasise that for the scope of the present article this field is needed at the intermediate step only, and by the end of the day it will be replaced by the constant $\mathrm{M}_R$.

Upon the local Weyl transformation
\begin{equation}
g_{\mu\nu} \rightarrow e^{2\phi} g_{\mu\nu},\quad H\rightarrow e^{-\phi} H,  \quad
\Delta_{u,d,L} \rightarrow e^{-\phi} \Delta_{u,d,L}, \quad S \rightarrow e^{-\phi}S, \quad \Omega \rightarrow e^{-\phi} \Omega,
\quad {\sigma \rightarrow e^{-\phi} \sigma} \label{WT}
\end{equation}
where $\phi$ is an arbitrary function of $x$, the ``intermediate" Dirac operator $\tilde{\mathcal{D}}$ transforms in a homogeneous way: 
\be
\tilde{\mathcal{D}} \longrightarrow e^{-\frac{5\phi}{2}} \tilde{\mathcal{D}}e^{\frac{3\phi}{2}},
\ee
and one can easily check (using the method of conformal variations, see for example~\cite{manual}) that the fourth heat kernel coefficient which is associated with $\tilde{\slashed{D}}^2$ 
is Weyl invariant. 

On the other side all heat kernel coefficients are gauge invariant. 
The only Weyl and gauge invariant combination 
of scalar fields of the dimension four which involves the  derivatives is\footnote{The trace is needed since we are dealing with the matrix valued scalar fields like $S$.}:
\begin{equation}
\mathrm{tr}\,\left[D_{\mu}({\rm scalar\,field})^{\dagger}D^{\mu}({\rm scalar\,field})\right] - \frac{1}{6} R\,
 \mathrm{tr}\,\left[({\rm scalar\,field})^{\dagger}({\rm scalar\,field}) \right], 
\label{Wisa}
\end{equation}
thus it is sufficient to compute the coefficient in front of $R ({\rm scalar\,field})^{\dagger}({\rm scalar\,field})$, whilst
the kinetic term, which contains all the covariant derivatives $D_{\mu}$, can be restored from \eqref{Wisa}. Note that for such a computation it is sufficient to consider constant scalar fields: $\partial_{\mu}(\mbox{scalar}) = 0$ and set the gauge connection  to zero. Since the same simplification is applicable for the $a_2$ contribution, let us assume it for a while. 
Using the well known Lichnerowics formula one can easily check that in our ``simplified" regime the endomorphism $E$, which enters in \eqref{a2} and \eqref{a4cg} equals to:
\be
E  = - M^2\otimes 1^{\rm s}_4 + \frac{R}{4}\otimes1_{384} \label{Esimp},
\ee
hence
\be
E^2 = M^4\otimes 1^{\rm s}_4 + \frac{R^2}{16}\otimes1_{384} - \left(\frac{R}{2}\right)\cdot M^2 \otimes 1^{\rm s}_4 \label{E2simp}
\ee
so the calculation of the new terms of the bosonic spectral action reduced to an algebraic exercise: one has to calculate $\mathrm{tr}\,M^2$ and $\mathrm{tr}\,M^4$. We remind that all the terms which disappeared because of our simplification can be recovered via the Weyl and the gauge invariance of $a_4$.

\subsection{Relevant traces}
One can check by a direct computation using e.g.\ Maple, the following formulas:
\begin{eqnarray}
{\rm tr}\, M^2  &=& {  2y_{1}}\sigma^2 + 4y_{2}{\left( \Omega^*\Omega\right)}
+ 4y_{3}{\left(\Delta_u^{\dagger}\Delta_u\right)} \nonumber\\
&+& 4y_{4}{\left(\Delta_d^{\dagger}\Delta_d\right)}
      +8y_{5}{\left(\Delta_L^{\dagger}\Delta_L\right)}+12 y_{6} {\left(H^{\dagger}H\right)}+4y_{7}\, {{\rm tr}\left(S^{\dagger}S\right)}, \label{M2}
\end{eqnarray}
and
\begin{eqnarray}
{\rm tr}\, M^4  &=&  2z_{1}\sigma^4  + 4z_2{\left(\Omega^*\Omega\right)^2}+4 z_{3} {\left(\Delta_u^{\dagger}\Delta_u\right)^2}
+ 4 z_{4}{\left(\Delta_d^{\dagger}\Delta_d\right)^2}
+8z_{5} {\left(\Delta_L^{\dagger}\Delta_L\right)^2} \nonumber\\
&& + 12 z_{6} {\left(H^{\dagger}H\right)^2} +8 z_{7}{\left(\Delta_u^{\dagger}\Delta_u\right)\left(\Omega^*\Omega\right) } 
+ 8z_{8} {\left(\Delta_d^{\dagger}\Delta_d\right)\left(\Omega^*\Omega\right) }
 \nonumber\\
&&+16z_{9}{\left(\Delta_u^{\dagger}\Delta_u\right)\left(H^{\dagger}H\right)} 
+16z_{10}{\left(\Delta_d^{\dagger}\Delta_d\right)\left(H^{\dagger}H\right)} 
+16z_{11}{\left(\Delta_L^{\dagger}\Delta_L\right)\left(H^{\dagger}H\right)}  \nonumber \\
&&+\left[ {8 z_{12}}{\left(\Delta_d^{\dagger}\Delta_L\right)\left(H^{\dagger}H\right)} + {8 z_{13}}{\left(\Delta_u^{\dagger}\Delta_L\right)\left(H^{\dagger}H\right)}
 + \mbox{c.c.}\right] \nonumber \\
 && 
+ 8z_{14}  {\left(\Omega^*\Omega\right)\left(H^{\dagger}H\right)} + 8z_{15} {\left(\Delta_u^{\dagger}\Delta_u\right)\sigma^2 } 
+ 8z_{16} {\left(\Omega^*\Omega\right)\sigma^2}  +  8z_{17} {\left(H^{\dagger}H\right) \sigma^2} \nonumber \\
&&
+ 4z_{18} \,{{\rm tr}\left(S^{\dagger} S\right)^2}  +{ 8z_{19} \left(H^{\dagger}H\right){\rm tr}\left(S^{\dagger}S\right) 
+ 8z_{20} \left(S\tilde{H}\right)^{\dagger}\left(S\tilde{H}\right) } \nonumber\\
&&{+8z_{21} \left(S{H}\right)^{\dagger}\left(S {H}\right)}
 + 8z_{22} {\left(\Delta_u^{\rm T}S\right)\left(\Delta_u^{\rm T}S\right)^{\dagger}}
 + 8z_{23} {\left(\Delta_L^{\dagger}\Delta_L\right)\,{\rm tr}\left(S^{\dagger}S\right)}  \nonumber \\
&& +\left[ 8 z_{24} {\left(\tilde{H}^{\rm T}S^{\rm T}\Delta_u\right)\sigma} 
 {+ 8 z_{25} \left(H^{\rm T} S^{\rm T}\Delta_u \Omega\right) } +\mbox{c.c}\right],  \label{M4}
\end{eqnarray}
where the constants $y_1$,...,$y_7$, $z_1$,...,$z_{25}$ depend on the Yukawa  couplings as follows: {
\begin{eqnarray} 
{y_{1}} & {\equiv}& { {\rm tr} \left(\hat y_M \hat y_M^{\dagger}\right)} \nonumber\\
y_{2} &\equiv& {\rm tr} \left(\hat y_{\Omega} \hat y_{\Omega}^{\dagger}\right) \nonumber\\
y_{3} &\equiv&  {\rm tr}\left(\hat y_{\Delta_u} \hat y_{\Delta_u}^{\dagger}\right)  \nonumber\\
y_{4} &\equiv&   {\rm tr} \left(\hat y_{\Delta_d} \hat y_{\Delta_d}^{\dagger}\right)  \nonumber\\
y_{5} &\equiv&   {\rm tr} \left(\hat y_{\Delta_L} \hat y_{\Delta_L}^{\dagger}\right)   \nonumber\\
y_{6} &{\equiv}& {{\rm tr} \left(\left[\hat y_u \hat y_u^{\dagger}\right] + \left[\hat y_d \hat y_d^{\dagger}\right]
+\frac{1}{3}\left[\hat Y_u \hat Y_u^{\dagger}\right] + \frac{1}{3}\left[\hat Y_d \hat Y_d^{\dagger}\right] \right)} \nonumber\\
y_{7} &\equiv& {\rm tr} \left(\hat y_S \hat y_S^{\dagger}\right) \nonumber\\
z_1 &{\equiv}&{ {\rm tr} \left(\hat y_M \hat y_M^{\dagger}\right)^2} \nonumber\\
z_2 &\equiv& {\rm tr} \left(\hat y_{\Omega} \hat y_{\Omega}^{\dagger}\right)^2 \nonumber\\
z_3 &\equiv&  {\rm tr}\left(\hat y_{\Delta_u} \hat y_{\Delta_u}^{\dagger}\right)^2  \nonumber\\
z_4 &\equiv&   {\rm tr} \left(\hat y_{\Delta_d} \hat y_{\Delta_d}^{\dagger}\right)^2  \nonumber\\
z_5 &\equiv&   {\rm tr} \left(\hat y_{\Delta_L} \hat y_{\Delta_L}^{\dagger}\right)^2   \nonumber\\
z_6 &{\equiv}&{ {\rm tr} \left(\left[\hat y_u \hat y_u^{\dagger}\right]^2 + \left[\hat y_d \hat y_d^{\dagger}\right]^2
+\frac{1}{3}\left[\hat Y_u \hat Y_u^{\dagger}\right]^2 + \frac{1}{3}\left[\hat Y_d \hat Y_d^{\dagger}\right]^2 \right)}\nonumber\\
z_7  &\equiv&{\rm tr}\left(\hat y_{\Omega} \hat y_{\Omega}^{\dagger}\right)
\left(\hat y_{\Delta_u}^{\dagger} \hat y_{\Delta_u}\right)\nonumber\\
z_8  &\equiv& {\rm tr}\left(\hat y_{\Omega}^{\dagger} \hat y_{\Omega}\right)
\left(\hat y_{\Delta_d}^{\rm T} \hat y_{\Delta_d}^*\right)\nonumber\\
z_9  &\equiv& \frac{1}{2} {\rm tr}\left[
\left(\hat y_{u} \hat y_{u}^{\dagger}\right)\left(\hat y_{\Delta_u}\hat y_{\Delta_u}^{\dagger}\right)  
+ \left(\hat Y_{u} \hat Y_{u}^{\dagger}\right)\left(\hat y_{\Delta_u}^{\dagger}\hat y_{\Delta_u}\right) 
\right]
\nonumber\\ 
z_{10}  &\equiv& \frac{1}{2} {\rm tr}\left[
\left(\hat y_{d} \hat y_{d}^{\dagger}\right)\left(\hat y_{\Delta_d}\hat y_{\Delta_d}^{\dagger}\right) 
+\left(\hat Y_{d} \hat Y_{d}^{\dagger}\right)\left(\hat y_{\Delta_d}^{\dagger}\hat y_{\Delta_d}\right) 
\right]
\nonumber\\ 
z_{11}  &\equiv& \frac{1}{2} {\rm tr}\left[
\left(\hat y_{u}^{\dagger} \hat y_{u}\right)\left(\hat y_{\Delta_L}\hat y_{\Delta_L}^{\dagger}\right) 
+  \left(\hat y_{d}^{\dagger} \hat y_{d}\right)\left(\hat y_{\Delta_L}\hat y_{\Delta_L}^{\dagger}\right)
+ \left(\hat Y_{u}^{\dagger} \hat Y_{u}\right)\left(\hat y_{\Delta_L}^{\dagger}\hat y_{\Delta_L}\right) 
+ \left(\hat Y_{d}^{\dagger} \hat Y_{d}\right)\left(\hat y_{\Delta_L}^{\dagger}\hat y_{\Delta_L}\right)
\right]
\nonumber\\ 
z_{12}  &\equiv& {\rm tr} \,\left(\hat y^{\dagger}_{\Delta_d}\hat y_d \hat y_{\Delta_L}\hat Y^{\dagger}_d\right)
\nonumber\\ 
z_{13}  &\equiv& {\rm tr} \,\left(\hat y^{\dagger}_{\Delta_u}\hat y_u \hat y_{\Delta_L}\hat Y^{\dagger}_u\right)
\nonumber\\ 
z_{14}  &\equiv& {\rm tr}\left[\left(
\hat y_{\Omega}\hat y^{\dagger}_{\Omega}\right)\left(\hat Y_u \hat Y^{\dagger}_u\right) 
+ \left(\hat y_{\Omega}^{\dagger}\hat y_{\Omega}\right)\left(\hat Y_d^* \hat Y^{\rm T}_d\right) \right] 
\nonumber\\
z_{15}  &\equiv& 
\frac{1}{2}{\rm tr}\left[\left(\hat y_M^{\dagger}y_M\right)\left(\hat y_{\Delta_u}^{\rm T}\hat y^*_{\Delta_u}\right)
+ \left(\hat y_M y_M^{\dagger}\right)\left(\hat y_{\Delta_u}^{\dagger}\hat y_{\Delta_u}\right)\right] \nonumber\\
z_{16}  &\equiv& \frac{1}{2}{\rm tr}\left[\left(\hat y_{\Omega}\hat y_{\Omega}^{\dagger}\right)\left(\hat y_{M}\hat y_{M}^{\dagger}\right)
+ \left(\hat y_{\Omega}^{*}\hat y_{\Omega}^{\rm T}\right)\left(\hat y_{M}^{\dagger}\hat y_{M}\right)\right] \nonumber\\
z_{17}  &{\equiv}&{ {\rm tr} \left(\hat y_M \hat y_M^{\dagger}\right)\left(\hat Y_u \hat Y_u^{\dagger}\right)}\nonumber\\
z_{18}  &\equiv& {\rm tr} \left(\hat y_S \hat y_S^{\dagger}\right)^2 \nonumber\\
{z_{19}  }&{\equiv}& {{\rm tr} \left[  \left(\hat y_S \hat y_S^{\dagger}\right)\left(\hat y_u^* \hat y_u^{\rm T} \right)
\right]}\nonumber\\
{z_{20}  }&{\equiv}& {{\rm tr} \left[ \left(\hat y_S^{\dagger} \hat y_S\right)\left(\hat Y_u^{\dagger} \hat Y_u \right)
\right]}\nonumber\\
{z_{21}  }&{\equiv}& {{\rm tr} \left[ \left(\hat y_S^{\dagger} \hat y_S\right)\left(\hat Y_d^{\dagger} \hat Y_d \right)
\right]}\nonumber\\
z_{22}  &\equiv& {\rm tr} \left[
 \left(\hat y_S\hat y_S^{\dagger} \right)\left(\hat y_{\Delta_u}^{*}\hat y_{\Delta_u}^{\rm T}\right)
\right] \nonumber\\
z_{23}  &\equiv& {\rm tr} \left[
\left(\hat y_S^{\dagger} \hat y_S\right)\left(\hat y_{\Delta_L}^{\dagger}\hat y_{\Delta_L}\right)
\right] \nonumber\\
z_{24}  &\equiv& {\rm tr} \left(\hat y_{\Delta_u}\hat y_M \hat Y_u^{*} \hat y_{S}^{\rm T} \right)
\nonumber \\
{z_{25} } &{\equiv}& {{\rm tr} \left(\hat{y}_{\Delta_y}\hat{y}_{\Omega}\hat{Y}_d^* \hat{y}_S^{\rm T} \right) }.
\end{eqnarray} 
}
\subsection{The full bosonic spectral action}
Substituting \eqref{M2}, \eqref{M4}, \eqref{Esimp} and \eqref{E2simp} in \eqref{a2} and \eqref{a4cg},  recovering the dependence on the derivatives and on the gauge fields according to \eqref{Wisa} and setting $\sigma = \mathrm{M}_R$ we arrive to the following answer for the new terms in the bosonic spectral action:  
\begin{eqnarray} 
&& S_B =S_B \big|_{\Delta_{u,d,S}=0,\Omega=0,S=0}  + \int d^4 x \sqrt{g^{\rm E}}\left\{- f_2\Lambda^2\left(\frac{y_{2}}{{\pi^2}}\Omega^*\Omega + \frac{y_{3}}{\pi^2}\Delta_u^{\dagger}\Delta_u
 \right. \right. \nonumber\\
&&\left.  
+ \frac{y_{4}}{\pi^2}\Delta_d^{\dagger}\Delta_d+ \frac{2y_{5}}{\pi^2}\Delta_L^{\dagger}\Delta_L
 + \frac{y_{7}}{{\pi^2}}{\rm tr \,} S^{\dagger}S  \right) 
+ f_4\left\{ \frac{y_{2}}{2\pi^2}\left(D_{\mu}\Omega^{*}D^{\mu}\Omega - \frac{R}{6}\Omega^{*}\Omega\right) \right.
\nonumber\\
&& \frac{ y_{3}}{2\pi^2}\left(D_{\mu}\Delta_u^{\dagger}D^{\mu}\Delta_u - \frac{R}{6}\Delta_u^{\dagger}\Delta_u\right) + \frac{ y_{4}}{2\pi^2}\left(D_{\mu}\Delta_d^{\dagger}D^{\mu}\Delta_d - \frac{R}{6}\Delta_d^{\dagger}\Delta_d\right)
 \nonumber\\
&& + \frac{ y_{5}}{\pi^2}\left(D_{\mu}\Delta_L^{\dagger}D^{\mu}\Delta_L - \frac{R}{6}\Delta_L^{\dagger}\Delta_L\right)
+\frac{y_{7}}{2\pi^2}{\rm tr \,}\left(D_{\mu}S^{\dagger}D^{\mu}S - \frac{R}{6}S^{\dagger}S\right)  
\nonumber\\
&& + \frac{1}{2\pi^2}z_2\left( \Omega^*\Omega\right)^2
 +\frac{1}{2\pi^2} z_{3} \left(\Delta_u^{\dagger}\Delta_u\right)^2
+\frac{1}{2\pi^2} z_{4}\left(\Delta_d^{\dagger}\Delta_d\right)^2
\nonumber\\
&& +\frac{1}{\pi^2}z_{5} \left(\Delta_L^{\dagger}\Delta_L\right)^2
+\frac{1}{\pi^2} z_{7} \left(\Delta_u^{\dagger}\Delta_u\right)\left(\Omega^*\Omega\right)
+ \frac{1}{\pi^2}z_{8}\left(\Delta_d^{\dagger}\Delta_d\right)\left(\Omega^*\Omega\right) 
 \nonumber\\
&&+\frac{2}{\pi^2}z_{9}\left(\Delta_u^{\dagger}\Delta_u\right)\left(H^{\dagger}H\right) +\frac{2}{\pi^2}z_{10}\left(\Delta_d^{\dagger}\Delta_d\right)\left(H^{\dagger}H\right)
+\frac{2}{\pi^2}z_{11}\left(\Delta_L^{\dagger}\Delta_L\right)\left(H^{\dagger}H\right)
 \nonumber \\
&& +\frac{{1}}{\pi^2} \,\left[ \left. {z_{12} }
\left(\Delta_d^{\dagger}\Delta_L\right)\left(H^{\dagger}H\right)+{z_{13} }
\left(\Delta_u^{\dagger}\Delta_L\right)\right.\left(H^{\dagger}H\right) + {\mbox{c.c.}} \right] \nonumber\\
&&  + \frac{1}{\pi^2}z_{14}  \left(\Omega^*\Omega\right)\left(H^{\dagger}H\right)
+ \frac{1}{\pi^2}z_{15}\left(\Delta_u^{\dagger}\Delta_u\right)\mathrm{M}_R^2 
 + \frac{1}{\pi^2}z_{16} \left(\Omega^*\Omega\right)\mathrm{M}_R^2 \nonumber\\
 && + \frac{1}{2\pi^2}z_{18} \left[{\rm tr \,}\left(S^{\dagger}S \right)\right]^2  
+{ \frac{1}{\pi^2}z_{19} \left(H^{\dagger}H\right){\rm tr}\left(S^{\dagger}S\right) 
+ \frac{1}{\pi^2}z_{20} \left(S\tilde{H}\right)^{\dagger}\left(S\tilde{H}\right) } \nonumber\\
&&{+\frac{1}{\pi^2}z_{21} \left(S{H}\right)^{\dagger}\left(S {H}\right)} + \frac{1}{\pi^2}z_{22}  {\left(\Delta_u^{\rm T}S\right)\left(\Delta_u^{\rm T}S\right)^{\dagger}} + \frac{1}{\pi^2}z_{23}  \left(\Delta_L^{\dagger}\Delta_L\right)  {\rm tr \,}\left(S^{\dagger}S \right) \nonumber\\
&&\left. +\frac{{1}}{\pi^2}\left[  z_{24} {\left(\tilde{H}^{\rm T}S^{\rm T}\Delta_u\right)\mathrm{M}_R} 
 {+  z_{25} \left(H^{\rm T} S^{\rm T}\Delta_u \Omega\right) } +\mbox{c.c}\right]\right\}. \label{finalanswer}
\end{eqnarray}
This is the result of the spectral action computation with the new fields coming form the Clifford requirement grading.

\section{Towards the physical action. \label{physact}}
{In this section we discuss how to make our spectral action applicable in  a physical context. 
In order to do this one has to carry out two important steps:
\begin{itemize}
\item{get rid of the redundant fermionic degrees of freedom,}
\item{make the action Lorentzian.}
\end{itemize}
The redundancy is usually solved projecting out the extra degrees of freedom~\cite{LMMS, Pepeetal, AC2M2}, while the Euclidean vs. Lorentzian issue has several ramifications (see for example~\cite{Dungen, Franco, DevFarLizMar, BiziBrouderBesnard}), but the usual method is to perform a Wick rotation.  We have shown in~\cite{direstraits} that the two issues are intimately related, and given a prescription on how to deal with them.

\subsection{General prescription: a review and discussion.}
Now we briefly recall how the Wick rotation works following \cite{direstraits}.
In order to pass from the Euclidean to a Lorentzian theory, each expression $F$ which involves the vierbeins $ e_{\mu}^a$ has to be transformed according to the following rule:
\be
\mbox{Wick:} \quad F\left[e_{\mu}^0, e_{\mu}^j \right]
\longrightarrow \quad F\left[\ii e_{\mu}^0, e_{\mu}^j\right],
\,\, \, j = 1,2,3. 
\label{bosWickvier}
\ee
As it was demonstrated in \cite{direstraits}, upon the transformation \eqref{bosWickvier} the Euclidean bosonic action $S_{\mathrm{bos}}^{\rm E}$, which comes out from the first three nonzero heat kernel coefficients, perfectly transforms into the ``textbook" Lorentzian action $S_{\mathrm{bos}}^{\rm M}$, in particular
\begin{equation}
\mbox{Wick:}\quad \exp{\left(-S_{\mathrm{bos}}^{\rm E}[{\rm fields}, g_{\mu\nu}^{\rm E}] \right)} 
\longrightarrow \exp{\left(\ii S_{\mathrm{bos}}^{\rm M}[{\rm fields}, g_{\mu\nu}^{\rm M}] \right)} \;, \label{cond}
\end{equation}
where the metric tensors $g_{\mu\nu}^{\rm E}$ and $g_{\mu\nu}^{\rm M}$ have the signatures $\left\{+,+,+,+\right\}$
 and $\left\{+,-,-,-\right\}$ respectively. We refer the reader to the quoted reference for the details.
 
A treatment of the fermionic action is more subtle, since the product space $\mathcal{H}$ contains extra degrees of freedom. Now we briefly recall what the problem is. The Hilbert space $\mathcal{H}$ of the almost commutative geometry has the following structure:
\be
\mathcal{H} = \mathrm{sp}(M)\otimes \mathcal{H}_F = \mathcal{H}_L \oplus \mathcal{H}_R \oplus \mathcal{H}_L^c \oplus \mathcal{H}_R^c, \label{Hprod}
\ee
where the subspaces $\mathcal{H}_L$, $\mathcal{H}_R$,  \emph{by definition} consist of the multiplets of the \emph{nonchiral} 4-component spinors
which transform under the gauge transformations \emph{as} the multiplets of the left handed and right handed \emph{chiral} fermions of the Standard Model, whilst the subspaces $\mathcal{H}_L^c$ and $\mathcal{H}_R^c$ consist of the \emph{nonchiral} 4-component spinors fermions which transform under the gauge transformations as the charge conjugated multiplets of the \emph{chiral}  left and  right fermions of the Standard Model. This doubling of the degrees of freedom is called in \cite{direstraits} the ``mirror doubling". On the one side the action of the Standard Model does not contain any independent variables with the index ``c", which indicates the charge conjugated field:  the charge conjugated spinor is obtained from the original one via the the charge conjugation operation (i.e. they are not independent variables, see \eqref{CM} below). This other doubling is called in \cite{direstraits} the ``charge conjugation doubling".

In order to get rid of the mirror doubling one has to extract the particles with the correct chirality. The left 
$\psi_{\mathcal{L}}$ and the right $\psi_{\mathcal{R}}$ chiral spinors are by definition the eigenstates of the left and the right chiral projectors:
\be
\psi_{\mathcal{L}} = \frac{1}{2}\left(1-\gamma^5\right)\psi_{\mathcal{L}}, \quad\quad
\psi_{\mathcal{R}} = \frac{1}{2}\left(1+\gamma^5\right)\psi_{\mathcal{R}}.
\ee
In order to get rid of the redundant fermionic degrees of freedom with the wrong chirality one has to extract just left chiral fermions from $\mathcal{H}_L$ and $\mathcal{H}_R^c$ and just right chiral fermions from $\mathcal{H}_R$ and $\mathcal{H}_L^c$, where we took into account  the fact that 
for the physical fermions, which live in Lorentzian space-time, the antiparticles have the opposite chirality with respect to the original particles.  So the subspace $\mathcal{H}_{+}$ of $\mathcal{H}$ which contains just the fermions with correct chiralities has the following structure:
\be
\mathcal{H}_{+}  = \left(H_L\right)_{\mathcal{L}} 
\oplus \left(\mathcal{H}_R\right)_{\mathcal{R}} \oplus \left(\mathcal{H}_L^c\right)_{\mathcal{R}} \oplus \left( \mathcal{H}_R^c\right)_{\mathcal{L}}. \label{Hp}
\ee
In the original paper \cite{AC2M2} such an extraction was presented in the form
\be
P_+ \mathcal{H}_{+} = \mathcal{H}_{+}
\ee
 where the  projector $P_+$ is defined via the grading as follows:
\be
P_{+} = \frac{1}{2}\left(1 + \gamma^5\otimes\gamma_F^{\mathrm{st}}\right). 
\ee
In this formula  $\gamma_F^{\mathrm{st}}$ stands for the ``standard" grading introduced in \cite{AC2M2}. Since we are working with the different grading $\gamma_F$, in order to arrive to the correct subspace \eqref{Hp} the connection between the projector 
$P_+$ and the grading $\gamma_F$ takes a slightly different form:
\be
P_{+} = \frac{1}{2}\left(1 + \gamma^5\otimes(\mathrm{Q} - \mathrm{L})\gamma_F\right).
\ee
The Euclidean fermionic action introduced in \cite{AC2M2}, which is free of the mirror doubling reads:
\be
{S}_F^{\mathrm{E}} = \frac{1}{2}\int d^4x \sqrt{g^{\mathrm{E}}}\left(\mathcal{J}\Psi_{+}\right)^{\dagger}\mathcal{D}\Psi_{+}, \quad \Psi_+ \in \mathcal{H}_+. \label{SFE}
\ee
As \cite{direstraits} shows, after the Wick rotation of the vierbeins \eqref{bosWickvier} one obtains:
\be
\mathrm{Wick}:\quad \exp\left( -S_F^{\mathrm{E}}[\mathrm{spinors}, e^a_{\mu}]\right) 
 \longrightarrow \exp\left(\ii{S}_F^{\mathrm{M\,\, doubled}}[\mathrm{spinors}, e^a_{\mu}]\right),
\ee
where the ``intermediate" fermionic action ${S}_F^{\mathrm{M\,\, doubled}}$ is already Lorentz invariant. However, due to the charge conjugation doubling, it depends on twice more fermionic fields than it is needed, it is not real and therefore it is not suitable for the canonical quantisation. In order to complete a construction of the physical fermionic action one has to eliminate the charge conjugation doubling  via
the following identification of the variables in the action ${S}_F^{\mathrm{M\,\, doubled}}$ from the subspaces $\mathcal{H}_L^c$ and $\mathcal{H}_R^{c}$ with the variables from $\mathcal{H}_L$ and $\mathcal{H}_R$:
\be
\mathrm{step \,\,1}: 
\begin{cases}
\;\left(\psi_L^{c}\right)_{\mathcal R} \in  \underbrace{ \left(H_L^c \right)_{\mathcal R}  }_{\subset H_+}
\quad\mbox{has to be identified with}\quad
 C_{\mathrm M}\left(\psi_L\right)_{\mathcal L} , \quad \left(\psi_L\right)_{\mathcal L}\in\underbrace{\left(H_L\right)_{\mathcal L}}_{\subset H_+} \\
\;\left(\psi_R^{c}\right)_{\mathcal L} \in  \smash{\underbrace{ \left(H_R^c \right)_{\mathcal L}  }_{\subset H_+}}
\quad\mbox{has to be identified with}\quad
 C_{\mathrm M}\left(\psi_R\right)_{\mathcal R} , \quad \left(\psi_R\right)_{\mathcal R}\in\smash{\underbrace{\left(H_R\right)_{\mathcal R}}_{\subset H_+}}
\rule{0pt}{16pt}
\end{cases} \label{fWick2},\vspace{14pt}
\ee
where the operation $C_{\mathrm M}$ is the charge conjugation operation, which acts on the arbitrary spinor $\psi$ as follows:
\be
C_{\mathrm M}\psi = -\ii \gamma^2\psi^*. \label{CM}
\ee
Note that in contrast to the Euclidean charge conjugation $J$ the operation $C_{\mathrm M}$ changes a chirality. We emphasise that the identification \eqref{fWick2} makes sense \emph{after} the Wick rotation to Lorentzian signature: since the quantities to be identified transform in the same way under the Lorentzian $SO(1,3)$ transformations rather than Euclidean $SO(4)$ rotations.
After the global axial transformation of all the remaining spinors
\be
\mathrm{step \,\,2}: \quad\psi \longrightarrow e^{-\frac{\ii\pi}{4}\gamma^{5}} \psi \label{axtrans}
\ee
one arrives to the ``textbook"  form ${S}_F^{\mathrm{M}}$ of the fermionic action:
\be
\mathrm{step \,\,1} + \mathrm{step \,\,2}:\quad \exp\left(\ii{S}_F^{\mathrm{M\,\, doubled}}[\mathrm{spinors}, e^a_{\mu}]\right) \longrightarrow \exp\left(\ii{S}_F^{\mathrm{M}}[\mathrm{phys. \,\, spinors}, e^a_{\mu}]\right).
\ee
Following \cite{direstraits} we remind that the last step must be performed before the quantisation: otherwise one will get an additional Pontrtyagin gauge action which comes out from the abelian axial anomaly. Below we apply these prescriptions to our model and we will find out a nontrivial outcome.

\subsection{This model}
Let us parametrise the elements of the Hilbert space $\mathcal{H}$ as follows:

\be
\Psi =  (\boldsymbol{\mathrm{v}}_R, \boldsymbol{\mathrm{e}}_R, \boldsymbol{\mathrm{L}}_L, 
           \boldsymbol{\mathrm{u}}_R, \boldsymbol{\mathrm{d}}_R, \boldsymbol{\mathrm{Q}}_L,   
           \boldsymbol{\mathrm{v}}_R^c, \boldsymbol{\mathrm{e}}_R^c, \boldsymbol{\mathrm{L}}_L^c,          
            \boldsymbol{\mathrm{u}}_R^c, \boldsymbol{\mathrm{d}}_R^c, \boldsymbol{\mathrm{Q}}_L^c             
             )^{\mathrm{T}},\label{HStruct}
\ee
This is basically the parametrisation \eqref{HfStruct} of the elements of $\mathcal{H}_F$, the change of typeface  indicates that the elements of $\mathcal{H}$ are spinors, no longer complex numbers. In these notations 
$\boldsymbol{\mathrm{u}}_R$ is a collection of 4-component spinors which transforms upon the action of the gauge group as the right handed quarks, $\boldsymbol{\mathrm{u}}_R^c$  is an \emph{independent} collection of  4-component spinors which transforms upon the action of the gauge group as the charge conjugated right handed quark field and so on.
The typical element of $\mathcal{H}_+$, which is constructed according to \eqref{Hp}, then becomes:
\be
\Psi_+ =(\left[\boldsymbol{\mathrm{u}}_R\right]_{\mathcal{R}}, \left[\boldsymbol{\mathrm{d}}_R\right]_{\mathcal{R}}, \left[\boldsymbol{\mathrm{Q}}_L\right]_{\mathcal{L}}, \left[\boldsymbol{\mathrm{v}}_R\right]_{\mathcal{R}}, \left[\boldsymbol{\mathrm{e}}_R\right]_{\mathcal{R}}, \left[\boldsymbol{\mathrm{L}}_L\right]_{\mathcal{L}}, \left[\boldsymbol{\mathrm{u}}^c_R\right]_{\mathcal{L}}, \left[\boldsymbol{\mathrm{d}}^c_R\right]_{\mathcal{L}}, \left[\boldsymbol{\mathrm{Q}}^c_L\right]_{\mathcal{R}}, \left[\boldsymbol{\mathrm{v}}^c_R\right]_{\mathcal{L}}, \left[\boldsymbol{\mathrm{e}}^c_R\right]_{\mathcal{L}}, \left[\boldsymbol{\mathrm{L}}^c_L\right]_{\mathcal{R}})^{\mathrm{T}}
\label{HStructP}
\ee
Composing the fermionic action \eqref{SFE}, applying the Wick rotation procedure \eqref{bosWickvier}, removing the charge conjugation doubling according to the general prescription \eqref{fWick2} and finally carrying out the axial transformation \eqref{axtrans} we see that:
\be
\mbox{Wick rotation + dequadrupling}: \quad -S_F^{\mathrm{E}} \longrightarrow \ii S^{\rm {M}}_F ,
\ee
where $S^{\rm {M}}_F $ is given by:
\begin{eqnarray}
 S^{\rm {M}}_F &=& \int d^4 x \sqrt{-g^{\rm M}} \left\{ \right.
 i  \overline{\left(\boldsymbol{\mathrm{u}}_{\mathcal R}\right)} \slashed{\nabla}\boldsymbol{\mathrm{u}}_{\mathcal R}
+ i \overline{\left(\boldsymbol{\mathrm{d}}_{\mathcal R}\right)} \slashed{\nabla} \boldsymbol{\mathrm{d}}_{\mathcal R}
+  i \overline{\left(\boldsymbol{\mathrm{Q}}_{\mathcal L}\right)} \slashed{\nabla} \boldsymbol{\mathrm{Q}}_{\mathcal L} \nonumber\\
&+& i  \overline{\left(\boldsymbol{\mathrm{v}}_{\mathcal R}\right)} \slashed{\nabla}\boldsymbol{\mathrm{v}}_{\mathcal R}
+ i \overline{\left(\boldsymbol{\mathrm{e}}_{\mathcal R}\right)} \slashed{\nabla} \boldsymbol{\mathrm{e}}_{\mathcal R}
+  i \overline{\left(\boldsymbol{\mathrm{L}}_{\mathcal L}\right)} \slashed{\nabla} \boldsymbol{\mathrm{L}}_{\mathcal L}  \nonumber \\
 &-& \left[ \overline{\left(\boldsymbol{\mathrm{Q}}_{\mathcal L}\right)} \left[\hat{y}^{\dagger}_u \otimes \tilde H\right] \boldsymbol{\mathrm{u}}_{\mathcal R}
+\overline{\left(\boldsymbol{\mathrm{Q}}_{\mathcal L}\right)}\left[\hat{y}_d^{\dagger} \otimes  H\right] \boldsymbol{\mathrm{d}}_{\mathcal R} + \right.
\nonumber\\
 &+& 
 \overline{\left(\boldsymbol{\mathrm{L}}_{\mathcal L}\right)}\left[\hat{Y}^{\dagger}_u \otimes \tilde H \right] \boldsymbol{\mathrm{v}}_{\mathcal R}
+ \overline{\left(\boldsymbol{\mathrm{L}}_{\mathcal L}\right)}\left[\hat{Y}^{\dagger}_d \otimes \ H\right] \boldsymbol{\mathrm{e}}_{\mathcal R} \nonumber\\
&+&\frac{1}{2} \overline{\left( C_{\rm M}\boldsymbol{\mathrm{v}}_{\mathcal R}\right)}\left[ \hat y_M^{\dagger}  \right]\boldsymbol{\mathrm{v}}_{\mathcal R}    
 \left.\left.
+ \overline{\left( C_{\rm M}\boldsymbol{\mathrm{e}}_{\mathcal R}\right)}\left[\hat y_{\Omega}^{\dagger}\otimes\Omega^*\right]\boldsymbol{\mathrm{v}}_{\mathcal R}  + \mbox{c.c.}\right]\right\}
 \label{SMink}.
\end{eqnarray}
In this formula for an arbitrary spinor $\psi$ the bar stands for the Dirac conjugation: $\bar\psi \equiv \psi^{\dagger} \gamma^0$.
Note that after the Wick rotation accompanied by the elimination of the fermionic quadrupling just the multiplets of the sructures 
$[\psi_R]_{\mathcal{R}}$ and $[\psi_L]_{\mathcal{L}}$ remain in the result, therefore we simplified the notations replacing them by
$\psi_{\mathcal{R}}$ and $\psi_{\mathcal{L}}$ respectively.

We now come to an important point of  this noncommutative geometric construction. Note that the fields $\Delta_u$, $\Delta_d$,  $\Delta_L$ and $S$, which are present in the Dirac operator and hence in the bosonic spectral action are absent in the fermionic action \eqref{SMink}! Let us clarify what has happened. If one looks carefully at the structure $\left(\mathcal{J}\Psi\right)^{\dagger} \mathcal{D} \Psi$, $\Psi\in \mathcal{H}$ one immediately
finds out that the mentioned fields always appear in interactions terms in the action (vertices) which involve  spinors with unphysical chiralities. \emph{Therefore, when one restricts the fermions just to the ``good-chirality" subspace $\mathcal{H}_+$, all these terms vanish}. Indeed, the $S$ field enters in $\left(\mathcal{J}\Psi\right)^{\dagger} \mathcal{D} \Psi$ in particular via the combination:
\bea
\left(J\boldsymbol{\mathrm{u}}_R\right)^{\dagger} \left(\gamma^5\otimes\hat{y}_S\otimes S \right)\boldsymbol{\mathrm{L}}_L   
\eea
When one restricts $\Psi\in H_{+}$ this expression turns into
\bea
\left(J\left[\boldsymbol{\mathrm{u}}_R\right]_{\mathcal{R}}\right)^{\dagger} \left(\gamma^5\otimes\hat{y}_S\otimes S \right)\left[\boldsymbol{\mathrm{L}}_L\right]_{\mathcal{L}}   &\equiv& \left(J P_{\mathcal{R}}\boldsymbol{\mathrm{u}}_R\right)^{\dagger} \left(\gamma^5\otimes\hat{y}_S\otimes S \right)P_{\mathcal{L}}\boldsymbol{\mathrm{L}}_L  
 \nonumber \\
&=& \left(J \boldsymbol{\mathrm{u}}_R\right)^{\dagger} \left(\gamma^5\otimes\hat{y}_S\otimes S \right)\left(P_{\mathcal{R}}P_{\mathcal{L}}\right)\boldsymbol{\mathrm{L}}_L    = 0,
\eea
where we took into account the fact that the chiral projectors $P_{\mathcal{L}} = \frac{1}{2}\left(1_4 - \gamma^5\right)$ and $P_{\mathcal{R}} = \frac{1}{2}\left(1_4 + \gamma^5\right)$ commute with the Euclidean charge conjugation $J$ and remain unchanged upon the Hermitian conjugation of matrices. One can easily check that all other combinations which involve $S$ vanish according to the same mechanism.

Now let us see what has happened to the $\Delta_{u,d,L}$ fields. The $\Delta_u$ field enters in $\left(\mathcal{J}\Psi\right)^{\dagger} \mathcal{D} \Psi$ in particular through the expression:
\be
\left(J\boldsymbol{\mathrm{u}}_R^c\right)^{\dagger} \left(\gamma^5\otimes\hat{y}_{\Delta_u}\otimes \Delta_u \right)\boldsymbol{\mathrm{v}}_R  
\ee
Upon the restriction $\Psi\in \mathcal{H}_+$ this expression turns into
\bea
\left(J\left[\boldsymbol{\mathrm{u}}_R^c\right]_{\mathcal{L}}\right)^{\dagger} \left(\gamma^5\otimes\hat{y}_{\Delta_u}\otimes \Delta_u \right)\left[\boldsymbol{\mathrm{v}}_R\right]_{\mathcal{R}}  &=& \left(J P_{\mathcal{L}}\boldsymbol{\mathrm{u}}_R^c\right)^{\dagger} \left(\gamma^5\otimes\hat{y}_{\Delta_u}\otimes \Delta_u \right)P_{\mathcal{R}}\boldsymbol{\mathrm{v}}_R  \nonumber\\
& =& \left(J \boldsymbol{\mathrm{u}}_R^c\right)^{\dagger} \left(\gamma^5\otimes\hat{y}_{\Delta_u}\otimes \Delta_u \right)P_{\mathcal{L}}P_{\mathcal{R}}\boldsymbol{\mathrm{v}}_R  = 0.
\eea 
One can easily check that all other terms, which involve the $\Delta$-fields vanish as well in the similar way.

\section{Conclusions and Outlook.}

Based on the purely algebraic idea to incorporate the Clifford structure in the finite dimensional spectral triple, proposed in \cite{FrancescoLudwik}, we arrive to a set of new scalar fields in the \emph{ minimal} version of the noncommutative standard model. Some of the new scalar fields (viz $\Delta_{u,d,L}$ and $S$) carry both colour and the weak isospin indexes. The fields of such a kind are of interest of recent phenomenological research, in particular the scalar lepto-quarks  are the case~\cite{leptoquarks}.

We computed the new terms in the bosonic spectral action which come out from these fields. The equation \eqref{finalanswer}
is one of the main results of this article.
The scalar-scalar couplings between the new fields and the Higgs field may improve the minimal noncommutative standard model from the phenomenological point of view: 
they  give positive contributions to the beta function of the Higgs self-interaction quartic constant at the level of the one loop~\cite{Machacek}, what is needed to avoid the 
vacuum instability problem~\cite{EliasMiro:2011aa,Bezrukov:2012sa}. 

We did not discuss in detail  in this paper the possible phenomenological consequences of these new terms. The whole approach to the standard model based on noncommutative geometry is now reaching the level to be confronted with phenomenology, and of course the scalar sector seems to be of paramount importance. The new fields discussed here may possibly be part of this, but more work is necessary in this direction.

The approach is interesting from the mathematical point of view as well.
It turns out that some of the new fields (viz $\Delta_{u,d,L}$ and $S$) are coupled to the spurious fermionic degrees of freedom, whose presence is due to the ``product-based" construction of the almost commutative spectral triple. Therefore  this model exhibits a very peculiar property, which is the another important result of this article. On the one side  these fields do not enter in the physical (Minkowskian) fermionic Lagrangian \eqref{SMink}, even though they appear in the Euclidean NCG Dirac operator. On the other side the physical (``Wick rotated" to the Lorentzian signature) bosonic spectral action keeps memory about these extra degrees of freedom: it depends on $\Delta_{u,d,L}$ and $S$. Therefore the fermionic quadrupling in the spectral approach is not just the  presence of the extra fermions to be projected out: by the end of the day it effects nontrivially the bosonic action of the model, without altering the fermionic action.

In this article we considered  an evolution of the spectral approach from an algebraic point of view.  There are other interesting mathematical directions which can be taken. In particular  we consider the manifold $\mathcal{M}$ without boundary, manifolds with boundaries have been considered within the spectral action formalism as well~\cite{ConnesChamseddineboundary1,ConnesChamseddineboundary2,Vassilevichboundary}. 
Recently another purely spectral feature has been discovered: parity anomaly on four dimensional manifolds with boundaries \cite{KurkovVassilevichPRD,KurkovVassilevichinpreparation}.  It would be interesting to understand the role played by the parity anomaly in the context of the spectral action approach, and the issue will deserve further scrutiny.


\subsection*{Acknowledgments}
We thank Francesco D'Andrea for several illuminating discussions which helped shape this paper, and clarifications of various conceptual and technical aspects of the Clifford-based approach.  
FL acknowledges the support of the COST action QSPACE, the INFN Iniziativa Specifica GeoSymQFT, the Spanish MINECO under project MDM-2014-0369 of ICCUB (Unidad de Excelencia `Maria de Maeztu'). M.K. acknowledges the support of the Sao-Paulo Research Foundation (FAPESP),
grant  2015/05120-0.

\end{document}